\documentclass[aps,prd,onecolumn,groupedaddress,nofootinbib,amssymb,nopacs
]{revtex4}
\usepackage[dvips]{graphicx}
\usepackage{amssymb}
\usepackage{amsmath}
\usepackage{graphicx,color}
\usepackage{amsfonts}
\usepackage{bm}
\usepackage{cancel}
\usepackage{comment}
\usepackage{timestamp}

\allowdisplaybreaks[4]
 
\usepackage{hyperref}

\hypersetup{
linktocpage=true,
colorlinks=true,
linkcolor=blue,          
citecolor=mygreen,        
filecolor=blue,      
urlcolor=blue           
}  
\usepackage{color}
\definecolor{mygreen}{rgb}{0.0,0.75,0.0}\usepackage{timestamp}

\usepackage{etoolbox}
\appto\UrlBreaks{\do\-}
\PassOptionsToPackage{hyphens}{url}
\usepackage[hyphenbreaks]{breakurl} 

\begin{document}

\tolerance=5000

\title{An approach to cold dark matter deviation and the $H_{0}$ tension problem \\by using machine learning}

\author{
Emilio Elizalde$^{1,2}$\footnote{E-mail: elizalde@ieec.uab.es}, 
Janusz Gluza$^{3}$\footnote{E-mail: janusz.gluza@us.edu.pl}, 
Martiros Khurshudyan$^{1,2,3}$\footnote{E-mail: khurshudyan@ice.csic.es}
}
\affiliation{
$^{1}$ Consejo Superior de Investigaciones Cient\'{\i}ficas, ICE/CSIC-IEEC,
Campus UAB, Carrer de Can Magrans s/n, 08193 Bellaterra (Barcelona) Spain \\
$^{2}$ International Laboratory for Theoretical Cosmology, Tomsk State University of Control Systems and Radioelectronics (TUSUR), 634050 Tomsk, Russia \\
$^{3}$ Institute of Physics, University of Silesia,  Katowice, Poland \\
}

\begin{abstract}
  
In this work, two different models, one with cosmological constant $\Lambda$, and baryonic and dark matter (with $\omega_{dm} \neq 0$), and the other with an $X$ dark energy (with $\omega_{de} \neq -1$), and baryonic and dark matter (with $\omega_{dm} \neq 0$), are investigated and compared. Using  Bayesian machine learning analysis, constraints on the free parameters of both models are obtained for the three redshift ranges: $z\in [0,2]$, $z\in [0,2.5]$, and $z\in [0,5]$, respectively.
For the first two redshift ranges, high-quality observations of the expansion rate $H(z)$ exist already, and they are used for validating the fitting results. Additionally, the extended range $z\in [0,5]$ provides predictions of the model parameters, verified when reliable higher-redshift $H(z)$ data are available. This learning procedure, based on the expansion rate data generated from the background dynamics of each model, shows that, at cosmological scales, there is a deviation from the cold dark matter paradigm, $\omega_{dm} \neq 0$, for all three redshift ranges. The results show that this approach may qualify as a solution to the $H_{0}$ tension problem. Indeed, it hints at how this issue could be effectively solved (or at least alleviated) in cosmological models with interacting dark energy.

\end{abstract}


\maketitle

\section{Introduction}\label{sec:INT}

Recently, an interesting issue which goes under the name of the $H_{0}$ tension problem has appeared. Specifically, the Planck-CMB data analysis provides a value $H_{0} = 67.4 \pm 0.5$ km/s/Mpc when the $\Lambda$CDM scenario is assumed \cite{Aghanim:2018eyx}.
But, a local measurement coming from the Hubble Space Telescope yields $H_{0} = 73.52 \pm 1.62$ km/s/Mpc \cite{Riess:2018byc}.
Thus, it has been pointed out that there is an actual tension between the calculated values of $H_{0}$ obtained from these two observational data sets, see also  \cite{Wong:2019kwg} and \cite{Freedman:2019jwv}. 
The difference between the two observed values of  $H_{0}$ is unexpectedly huge. Thus, on top of the other problems of modern cosmology, we have an additional one, which indicates that either there is new physics here or maybe a serious problem with either of the two measurements. Indeed, at this moment, it is very hard to estimate where the problem lies since we still need to reveal the nature of dark energy from one side, and the nature of dark matter, from the other. Anyhow, if the problem would eventually point towards new physics, then we would be in a perfect starting situation to clarify, constrain and select good dark energy models as extensions and modifications of General Relativity. However, if it turned out to be a measurement-related issue, then we would be dealing with an artificially crafted problem, not carrying useful information for cosmology. In any case, it is too early to reach a final conclusion, as has been discussed in the recent literature. It has been shown that some of the interacting dynamical dark energy models available on the market could be employed to solve the issue. But also, a modification of General Relativity may well lead to other solutions 
\cite{Sharov:2020bnk,Amirhashchi:2020qep,Braglia:2020iik,Wu:2020nxz,Alestas:2020mvb,Wang:2020zfv,Blinov:2020uvz,Tian:2020tur,Sakstein:2019fmf,vanPutten:2017qte,DiValentino:2019jae,DiValentino:2019ffd,DiValentino:2020naf,Cai:2019bdh,Briffa:2020qli,Yao:2020pji,Yang:2021flj,Nunes:2018xbm}
(to mention some references). 

The idea of considering interacting dark energy models is not new and has been used successfully in cosmology over many years. In interacting dark energy scenarios, one operates at the macroscopic level and introduces a non-gravitational interaction between dark energy and dark matter. This can be viewed as an artificial intervention and, to our knowledge, up to this moment, no well-defined theory has been found that can explain where the non-gravitational interaction comes from. One draws the existence of non-gravitational interactions between dark energy and dark matter merely from observations. However, we should recognize that this could be an artifact created  by a non-proper screening of our observational Universe. This could induce an improper bias, which added to the lack of a full understanding of the accelerated expansion of the Universe, could create the illusion of the additional interaction. However, we would like to stress that, in no way the above discussion wants to diminish the real value of the huge amount of work done in this direction, while exploring such possibility. It is just a motivation for additional research following a different path, since we do believe that there is yet job to be done in order to answer such questions. The community is still very active in exploring new and interesting aspects of interacting dark energy models~\cite{Bamba:2012cp,Odintsov:2017icc,Brevik:2017juz,Capozziello:2005pa,Capozziello:2018mds,Brevik:2019mah,Nojiri:2005sr,Brevik:2012ka,Brevik:2012nt,Brevik:2017msy,Mishra:2017zce,Odintsov:2018obx,Odintsov:2020voa,Yang:2020wby,Yang:2019nhz,Yang:2019fjt,Li:2019loh, Elizalde:2018ahd,Yang:2018qmz,Elizalde:2017dmu,Khurshudyan:2015mva,Aljaf:2020eqh,Aljaf:2019ilr,Khurshudyan:2018kfk,Khurshudyan:2016qox,Khurshudyan:2016zhu,Khurshudyan:2016xst,Khurshudyan:2015mpa,Khurshudyan:2017qtd,Khurshudyan:2017fxd,Sadri:2019yqs}
~(to mention some references). 

Before focusing on the main goal of our work, we would like to recall a way to infer cosmological knowledge from observational data. It is based on the well known Bayes theorem, which allows connecting the posterior distribution with the prior one with a likelihood. It has been applied very often to constrain the models yielding new knowledge. We shall omit technical details, but we would like to mention that we need to use a very well-defined sampling technique to build the posterior distribution. In the literature, there are interesting approaches mainly based on different ideas on how to deal with rejected and accepted samplings. The last ones significantly improve the statistics and speed up the fitting process. However, we need to have very well-defined models to proceed with the calculations efficiently.  

Retaking the above discussion on interacting dark energy models, let us now clarify what we mean by a well-defined model; namely, in the process, we need to supply an explicit form of dark energy, dark matter and the form of the non-gravitational interaction between them. Then, at the end of the day, constraining the model, we can make some inference. In particular, we can claim whether a specific type of non-gravitational interacting dark energy model is supported or not. However, there is another method known as Gaussian Process, which has been used in the recent literature very intensively~(see for instance \cite{Cai:2019bdh} 
and \cite{Briffa:2020qli} 
and references therein for more details). It is one of the Machine Learning approaches allowing to do reconstruction directly from data. In this case, for the same interacting dark energy models, we can reconstruct the possible non-gravitational interaction term as soon as the models of dark energy and dark matter are given. In this case, we do not need to supply any parametric form describing the interaction. 

In other words, following this approach, we can get the most general information about the specific task in a model independent way just having access to the observational data. The method has already gained a lot of attention, and we refer the reader to our reference list for more details on the topic. Huge efforts are being made attempting to integrate different Machine Learning algorithms into cosmology, and astrophysics \cite{Yan:2020wsr,Wang:2020sxl,Varma:2020kbq,Mangena:2020jdo,Escamilla-Rivera:2019hqt,Green:2020hst,Aljaf:2020nsl,Arjona:2020skf,Elizalde:2020mfs,Elizalde:2020pps,Elizalde:2020say}. 
There are different reasons for this, but the main one is the ability of different Machine Learning algorithms to learn the reason leading to the answer. One of them is the Bayesian (Probabilistic) Machine Learning approach, which, generally speaking, uses probabilities to think about probabilities. Again omitting all details and the history behind the approach, we would like to mention that it is beneficial to gain information about the model, which would be extremely hard to do with the usual approaches requiring data collection. This is because it is based on the generative process directly connected to the model, and, having it, we can explore the regions from which, in practice, it is tough to get answers. It should be mentioned that, recently, this method has been applied with success to study the $H_{0}$ tension problem in a single inhomogeneous viscous Universe. Moreover, it has been demonstrated that the $H_{0}$ tension problem solution is very weakly dependent on the specific form of the viscosity term. Instead, it has been shown that the solution mostly depends on the equation of state parameter of the barotropic fluid, used to craft the viscous dark energy fluid model \cite{Elizalde:2020mfs}. 
On the other hand, using  Bayesian (Probabilistic) Machine Learning, the cosmic opacity has been constrained, revealing its connection with the $H_{0}$ tension problem \cite{Elizalde:2020pps}. 
In another paper, we surprisingly found that the Swampland criteria itself can be used to constrain the models for the dark energy-dominated Universe \cite{Elizalde:2020say}. 
Motivated by the previously obtained results, we here continue to apply the same method to understand the extent to which we can use it and the potential limitations that one could face. 

This work aims to reveal possible connections between dark matter and the $H_{0}$ tension problem by using a Bayesian (Probabilistic) Machine Learning approach. The motivation comes from the ability of interacting dark energy models to solve the mentioned tension problem and how non-gravitational interactions may appear in cosmology. In particular, the presence of a non-gravitational interaction will not preserve the energy densities of each component separately anymore, which would force us to accept that, at the cosmological scales, we do not have purely cold dark matter. In this case, it is reasonable to assume that this deviation from cold dark matter can be a factor in solving (or alleviating at least) the $H_{0}$ tension problem. Motivated by the mentioned results, we would like to apply the Bayesian (Probabilistic) Machine Learning approach to study the $H_{0}$ tension problem and see how it is connected to the dark matter issue. It is easy to see that, with interacting dark energy models, the interaction affects the equation of state parameter of both components. It is then tough to separate the dark energy and dark matter contribution in the solution of the $H_{0}$ tension problem. 

The models we shall here study are one with cosmological constant $\Lambda$, and baryonic and dark matter (with $\omega_{dm} \neq 0$), and another with an $X$ dark energy (with $\omega_{de} \neq -1$), and baryonic and dark matter (with $\omega_{dm} \neq 0$), which are investigated and compared. Here, $\omega_{dm}$ is considered to be one of the free parameters. The constraints  are obtained for the three redshift ranges: $z\in [0,2]$, $z\in [0,2.5]$, and $z\in [0,5]$, respectively.
The first two redshift ranges are chosen to validate our fit results since we have very good quality observations of the expansion rate $H(z)$ in these cases. Additionally, the extended range $z\in [0,5]$ provides predictions of the model parameters, verified when reliable higher-redshift $H(z)$ data would become available. The learning method based on the generated $H(z)$ expansion rate data from the background dynamics of each model shows that, on cosmological scales, there is a deviation from the cold dark matter paradigm (it is not related to non-gravitational interaction). Moreover, we will encounter a hint that it could solve the tension problem and another on why the parameters of the interaction term should be small, as is discussed in the literature. Of course, to understand to which extend the observed deviation can be used to overcome the various cosmological problems, instead of involving interacting dark energy models, is still the issue, which will have to be tackled down in forthcoming papers. 

This paper is organized as follows. The philosophy behind the method used here is discussed in Sect.~\ref{sec:BIBBL}. In the same section, we present the models to be studied. The results are discussed in  Sect.~\ref{sec:BEM}, which is followed by an analysis of the consequences, both interesting for the $H_{0}$ tension and for the accelerated expanding Universe issues. The model-based generative process employed in the Bayesian Learning approach is built from Eqs.~(\ref{eq:HXCDM}) and~(\ref{eq:EXCDM}), by assuming that $\omega_{dm}$ is one of the free parameters of the model. The model-based generative process and the analysis leading to our final results are performed by using PyMC3 \cite{salvatier2015probabilistic}. Final conclusions of our analysis are given in Sect.~\ref{sec:conc}.

\section{The method and models}\label{sec:BIBBL}

Our analysis is based on the Bayesian (Probabilistic) Machine Learning approach, and in this section, we will briefly describe its key points. It is an interesting method that has been used to tackle various problems. Its advantage is due to the generative process being used, which allows exploring regions that are usually unreachable for some reason. Moreover, recent studies show that using it with reinforcement and inverse reinforcement can significantly speed up the learning process and, at the same time, reduce the costs of collecting data. On the other hand, it is clear that to build the generative process, we need to have a very good mathematical model describing the interesting processes for us. Because now only the generated data will be the key ingredient involved in this learning process. Fortunately, over the years, we have learned how to craft good models interesting for cosmology and astrophysics. Therefore, it is reasonable to apply the Bayesian (Probabilistic) Machine Learning approach to different problems from those fields and make some inference. 

In particular, it can be used for those problems, which cannot be fully explored by other means due to the low quality of data. Moreover, since we use a generative process, it is possible to do some forecasting, too, which can help design new missions and observations. In the recent literature, the basic ideas behind Bayesian Machine Learning have been well documented. Moreover, various frameworks allow the organization of Bayesian (Probabilistic) Machine Learning significantly, reducing the coding time, which allows concentrating on the physics of the process. One of such frameworks is the python-based PyMC3 framework, which can be easily integrated with other python-based frameworks, too \cite{salvatier2015probabilistic}
\footnote{We refer the readers to the nice documentation supplied with PyMC3, for more details on the Bayesian (Probabilistic) Machine Learning approach. Moreover, an existing set of interesting tasks and its solutions can generate intuition when and how Bayesian (Probabilistic) Machine Learning can be used.}. It should be noted that it uses Theano \cite{bergstra-proc-scipy-2010}. The last one is a deep learning python-based library to construct probability distributions and implement cutting edge inference algorithms.

Now, if we want to use Bayesian (Probabilistic) Machine Learning, we need to go through the following three steps:
 
\begin{itemize}

\item We need to define the model to be used to provide a so-called generative process. In our case, it will be the cosmological model.

\item We should envisage the data to be the data obtained from the generative process. In our case, it will be the expansion rate data generated from the cosmological model, which we are going to study\footnote{We direct the readers to 
\cite{Elizalde:2020say}-\cite{valelis2020studying} 
for more details about the possible difference between Machine Learning and Bayesian Machine Learning that rose namely in this step.}. 

\item We then run the learning algorithm to get a brand new distribution over the model parameters and update our prior belief. In other words, we need the prior distributions over the model's free parameters. This will correspond to our initial beliefs. Moreover, we need to update them each time to get the posteriors. In our case, we use PyMC3 and in-built functions to do all required steps.
\end{itemize}

Now let us discuss the models that we will study in our paper. In particular, we will consider two cosmological models having specific expansion rate history, obtained from a more general one presented below
\begin{equation}\label{eq:HXCDM}
H(z) = H_{0} E(z,\Omega_{dm}, \Omega_{b}, \omega_{dm}, \omega_{de}),
\end{equation}
with 
\begin{equation}\label{eq:EXCDM}
E(z,\Omega_{dm}, \Omega_{b}, \omega_{dm}, \omega_{de}) = \left [ \Omega_{b} (1+z)^{3} + \Omega_{dm} (1+z)^{3(1+\omega_{dm})}  +  (1- \Omega_{dm} - \Omega_{b}) (1+z)^{3(1+\omega_{de})}\right]^{1/2},
\end{equation}
where $H_{0}$, $\Omega_{b}$ and $\Omega_{dm}$ are the Hubble parameter, and the fraction of the baryonic matter and dark matter, at $z=0$, respectively. On the other hand, $\omega_{dm}$ and $\omega_{de}$ are the equation of state parameters describing dark matter and dark energy, respectively. Moreover, if we consider $\omega_{de} = -1$ and $\omega_{dm} = 0$, we reduce the expansion rate given by Eqs.~(\ref{eq:HXCDM}) and~(\ref{eq:EXCDM}) to the one corresponding to the standard $\Lambda$CDM model. While, if we consider $\omega_{de} \neq -1$, $\omega_{dm} = 0$ and combine dark matter with baryonic matter, we get the known XCDM model. 

Our goal is to constrain two cosmological models using the Bayesian (Probabilistic) Machine Learning approach, where the expansion rate data will be generated. The models are, namely, one with a cosmological constant $\Lambda$, and baryonic and dark matter (with $\omega_{dm} \neq 0$), and another with an $X$ dark energy (with $\omega_{de} \neq -1$), and baryonic and dark matter (with $\omega_{dm} \neq 0$). In both cases, $\omega_{dm}$ is a free parameter and will be constrained in three different redshift ranges, namely $z\in [0,2]$, $z\in [0,2.5]$, and $z\in [0,5]$, respectively. This strategy will allow us to estimate possible deviations of $\omega_{dm}$ from zero and explore how this is related to the $H_{0}$ tension problem. The first two redshift ranges are chosen to validate our fitting results since we have very good quality observations of the expansion rate $H(z)$ in these cases~(see Table~\ref{tab:Table0}). 

\begin{table}[t]
  \centering
    \begin{tabular}{ |  l   l   l  |  l   l  l  | p{2cm} |}
    \hline
$z$ & $H(z)$ & $\sigma_{H}$ & $z$ & $H(z)$ & $\sigma_{H}$ \\
      \hline
$0.070$ & $69$ & $19.6$ & $0.4783$ & $80.9$ & $9$ \\
         
$0.090$ & $69$ & $12$ & $0.480$ & $97$ & $62$ \\
    
$0.120$ & $68.6$ & $26.2$ &  $0.593$ & $104$ & $13$  \\
 
$0.170$ & $83$ & $8$ & $0.680$ & $92$ & $8$  \\
      
$0.179$ & $75$ & $4$ &  $0.781$ & $105$ & $12$ \\
       
$0.199$ & $75$ & $5$ &  $0.875$ & $125$ & $17$ \\
     
$0.200$ & $72.9$ & $29.6$ &  $0.880$ & $90$ & $40$ \\
     
$0.270$ & $77$ & $14$ &  $0.900$ & $117$ & $23$ \\
       
$0.280$ & $88.8$ & $36.6$ &  $1.037$ & $154$ & $20$ \\
      
$0.352$ & $83$ & $14$ & $1.300$ & $168$ & $17$ \\
       
$0.3802$ & $83$ & $13.5$ &  $1.363$ & $160$ & $33.6$ \\
      
$0.400$ & $95$ & $17$ & $1.4307$ & $177$ & $18$ \\

$0.4004$ & $77$ & $10.2$ & $1.530$ & $140$ & $14$ \\
     
$0.4247$ & $87.1$ & $11.1$ & $1.750$ & $202$ & $40$ \\
     
$0.44497$ & $92.8$ & $12.9$ & $1.965$ & $186.5$ & $50.4$ \\

$$ & $$ & $$ & $$ & $$ & $$\\ 

$0.24$ & $79.69$ & $2.65$ & $0.60$ & $87.9$ & $6.1$ \\
$0.35$ & $84.4$ & $7$ &  $0.73$ & $97.3$ & $7.0$ \\
$0.43$ & $86.45$ & $3.68$ &  $2.30$ & $224$ & $8$ \\
$0.44$ & $82.6$ & $7.8$ &  $2.34$ & $222$ & $7$ \\
$0.57$ & $92.4$ & $4.5$ &  $2.36$ & $226$ & $8$ \\ 
          \hline
    \end{tabular}
    \vspace{5mm}
\caption{$H(z)$ and its uncertainty $\sigma_{H}$ are in the units of km s$^{-1}$ Mpc$^{-1}$. The upper panel consists of thirty samples deduced from the differential age method. The lower panel corresponds to ten samples obtained from the radial BAO method. The table is according to \cite{Cai:2019bdh} 
~(see also references therein for details).}
  \label{tab:Table0}
\end{table}

Additionally, we consider the  $z\in [0,5]$ extended redshift range for forecasting reasons; the full validation of the results will have to wait for the near future when observations of higher redshift $H(z)$ data become available. As we consider the $H_{0}$ tension problem, then the expansion rate data to validate our results is the most obvious choice. The approach can be used to consider other observations, too. However, we wanted to avoid any bias that can be artificially induced from merging different data sets at this stage. 

Before ending this section, we should recall the standard notation of FLRW cosmology~(with $8 \pi G = c = 1$). In particular, the metric, in this case, has the form
\begin{equation}
ds^{2} = -dt^{2} + a(t)^{2} \sum_{i =1}^{3} (dx^{i})^{2},
\end{equation}
and  
\begin{equation}\label{eq:F1}
H^{2} = \frac{1}{3}\sum_{k}\rho_{k}.
\end{equation}
Moreover, the energy conservation law for each energy source can be expressed as
\begin{equation}\label{eq:EC}
\dot{\rho}_{k} + 3 H (\rho_{k} + P_{k}) = 0,
\end{equation}
with $P_{k} = \omega_{k} \rho_{k}$, where the index $k$ stands for baryons, dark matter and dark energy, respectively. Due to the mentioned assumptions, we end up on the background dynamics given by Eqs.~(\ref{eq:HXCDM}) and (\ref{eq:EXCDM}). It should be noted that Eq.~(\ref{eq:EC}) gives the evolution of the energy densities of the non-interacting fluids representing the energy budget of the Universe. However, if we consider for instance,  $Q$, a non-gravitational interaction between dark matter and dark energy only, then Eq.~(\ref{eq:EC}) turns out to be the following 
\begin{equation}\label{eq:EC_b}
\dot{\rho}_{b} + 3 H \rho_{b} = 0,
\end{equation}
\begin{equation}\label{eq:EC_dm}
\dot{\rho}_{dm} + 3 H (\rho_{dm} + P_{dm}) = Q,
\end{equation}
and
\begin{equation}\label{eq:EC_de}
\dot{\rho}_{de} + 3 H (\rho_{de} + P_{de}) = -Q.
\end{equation}
A closer look to Eqs.~(\ref{eq:EC_dm}) and (\ref{eq:EC_de}) indicates that non-gravitational interaction affects the equation of state parameters of both components. Even if we choose pure cold dark matter, $P_{dm} = 0$, the interaction $Q$ can make it colder or warmer, depending on its form and the constraints on its free parameters. The very brief discussion on interacting dark energy models given above will certainly help the reader follow our results. More details on interacting dark energy models can be found in the references of this paper.

\section{Results and discussion}\label{sec:BEM}

In this section, we present and discuss our results. We have divided it into two subsections, intending to simplify the presentation. It had been mentioned already that we consider three redshift ranges for both cosmological models, namely $z\in [0,2]$, $z\in [0,2.5]$ and $z \in [0,5]$, to understand better how our knowledge about the dark matter equation of state parameter can be changed. We remind the reader that our analysis is based on a generative process, where the generated data is the expansion rate of the Universe. This follows from the fact that we are primarily interested in exploring the connection between the $H_{0}$ tension problem and the nature of dark matter. In our opinion, this setup allows us to avoid possible artificial biases that could otherwise appear by using several datasets together, masking the true nature of the problem. Eventually, we can extend this analysis to include other types of "observational" data allowing us to understand the sources of the bias responsible for non-proper conclusions. We will undertake this specific analysis in another paper. The two tables, Table~\ref{tab:Table1} and Table~\ref{tab:Table2}, summarize our results here.

\subsection{Model with $\omega_{dm} \neq 0$ dark matter and cosmological constant}

The first example we have considered is a model of the Universe where the expansion rate is given by Eq.~(\ref{eq:HXCDM}) with
\begin{equation}\label{eq:EXCDM_M1}
E(z,\Omega_{dm}, \Omega_{b}, \omega_{dm}) = \left [ \Omega_{b} (1+z)^{3} + \Omega_{dm} (1+z)^{3(1+\omega_{dm})}  +  \Omega_{de} \right]^{1/2},
\end{equation}
where $\Omega_{de} = 1- \Omega_{dm} - \Omega_{b}$. It is a model of the Universe with the cosmological constant as its dark energy and baryonic and dark matter, with $\omega_{dm} \neq 0$. We now devise a generative process based on Bayesian (Probabilistic) Machine Learning to constrain $\omega_{dm}$ and estimate or predict (depending on the redshift range considered) the dynamics of changes that it can undergo based on the high redshift expansion rate measurements. The other parameters of the model to be fitted are $H_{0}$, $\Omega_{dm}$ and $\Omega_{b}$, respectively. On the other hand, as we discuss below, in this way, we can explore other reasons behind the $H_{0}$ tension problem. In particular, someones not related, say, to the dark energy problem. The constraints discussed below result from the learning process based on Eqs.~(\ref{eq:HXCDM}) and  (\ref{eq:EXCDM_M1}) using $H_{0} \in [64.0, 80.0]$,  $\Omega_{dm} \in [0.23, 0.4]$, $\Omega_{b} \in [0.0, 0.1]$ and $\omega_{dm} \in [-0.1, 0.1]$ as flat priors. In other words, we have generated data, using Eqs.~(\ref{eq:HXCDM}) and (\ref{eq:EXCDM_M1}), by imposing the mentioned priors. The results of the learning procedure can be read from Table~\ref{tab:Table1} and the ensuing presentation. 

\begin{figure}[h!]
 \begin{center}$
 \begin{array}{cccc}
\includegraphics[width=120 mm]{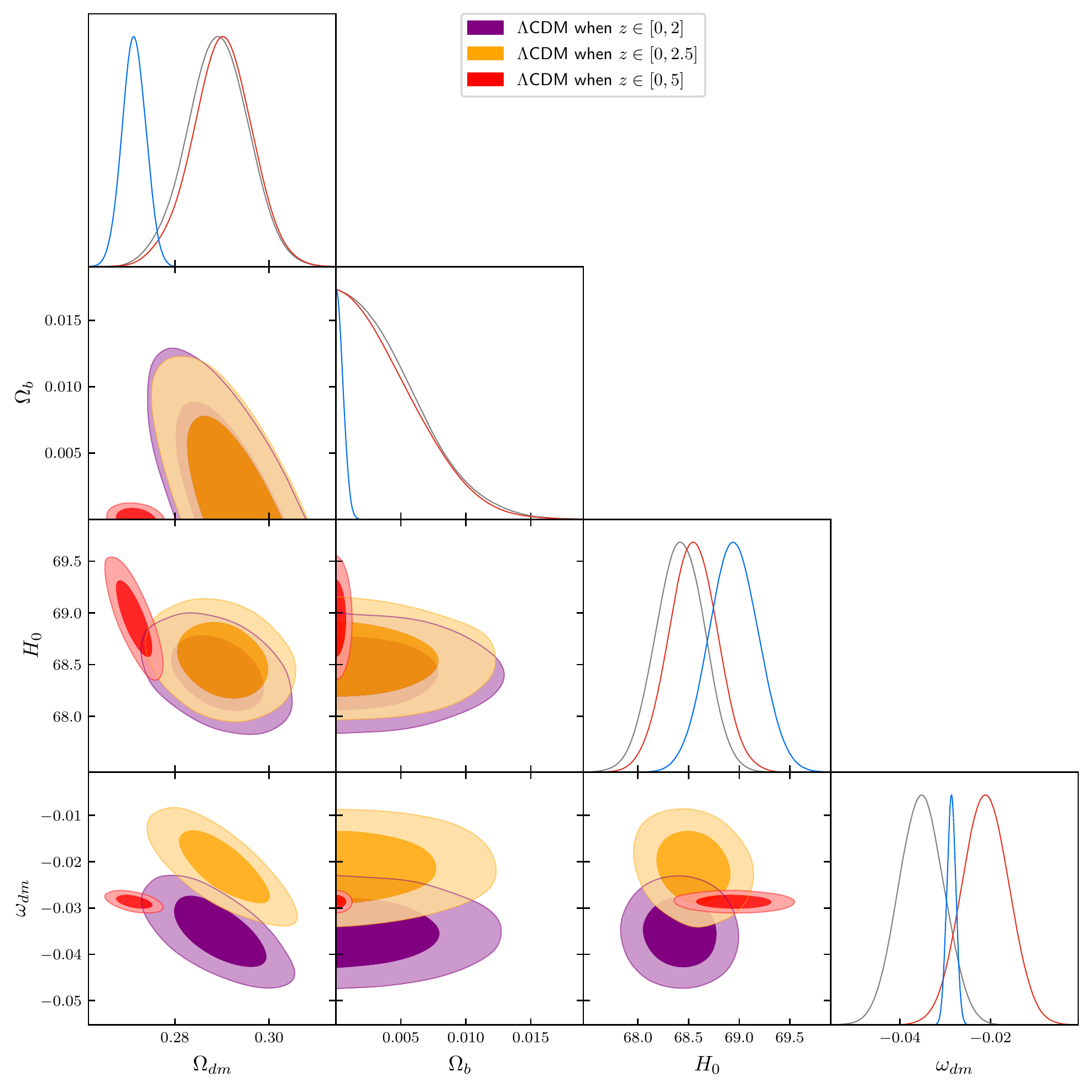}
 \end{array}$
 \end{center}
\caption{Contour maps of the model given by Eqs.~(\ref{eq:HXCDM}) and (\ref{eq:EXCDM_M1}) for $z \in [0,2]$, $z \in [0,2.5]$ and $z \in [0,5]$, respectively. The best fit values of the model parameters have been found to be $H_{0} = 68.42 \pm 0.24$ km/s/Mpc, $ \Omega_{dm} = 0.289 \pm 0.007$, $\Omega_{b} = 0.0037 ^{+0.0032}_{+ 0.0025}$ and $\omega_{dm} = -0.035 \pm 0.005$ when $z \in [0,2]$. While, when  $z \in [0,2.5]$, the best fit values of the model parameters have been found to be $H_{0} = 68.54 \pm 0.24$ km/s/Mpc,  $ \Omega_{dm} = 0.291 \pm 0.007$, $\Omega_{b} = 0.0035 ^{+ 0.0036}_{-0.0024}$ and $\omega_{dm} = -0.022 \pm 0.005$. Eventually, the best fit values of the model parameters have been found to be $H_{0} = 68.95 \pm 0.25$ km/s/Mpc,  $ \Omega_{dm} = 0.271 \pm 0.003$, $\Omega_{b} = 0.00035^{+ 0.00038}_{-0.00022}$ and $\omega_{dm} = -0.028 \pm 0.001$, when $z \in [0,5]$. In all three cases, $H_{0} \in [64.0, 80.0]$,  $\Omega_{dm} \in [0.23, 0.4]$, $\Omega_{b} \in [0.0, 0.1]$, and $\omega_{dm} \in [-0.1, 0.1]$ flat priors have been imposed during the generative process used to generate the "observational" data. The analysis is based on 10 chains and, in each chain, 10,000 "observational" data-sets from the model have been simulated or generated.}
 \label{fig:Fig0_1_a}
\end{figure}

\begin{itemize}

\item The best fit values of the model parameters are $H_{0} = 68.42 \pm 0.24$ km/s/Mpc, $ \Omega_{dm} = 0.289 \pm 0.007$, $\Omega_{b} = 0.0037 ^{+0.0032}_{+ 0.0025}$, and $\omega_{dm} = -0.035 \pm 0.005$, when $z \in [0,2]$. The contour map is given in Fig.~(\ref{fig:Fig0_1_a}), in purple colour. 

\item  The best fit values of the model parameters will most likely be $H_{0} = 68.54 \pm 0.24$ km/s/Mpc,  $ \Omega_{dm} = 0.291 \pm 0.007$, $\Omega_{b} = 0.0035 ^{+ 0.0036}_{-0.0024}$, and $\omega_{dm} = -0.022 \pm 0.005$, when $z \in [0,2.5]$. The contour map is given in Fig.~(\ref{fig:Fig0_1_a}), in orange colour.

\item  The best fit values of the model parameters will most likely be $H_{0} = 68.95 \pm 0.25$ km/s/Mpc,  $ \Omega_{dm} = 0.271 \pm 0.003$, $\Omega_{b} = 0.00035^{+ 0.00038}_{-0.00022}$, and $\omega_{dm} = -0.028 \pm 0.001$, when $z \in [0,5]$. The contour map, when $z \in [0, 5]$, is given in Fig.~(\ref{fig:Fig0_1_a}), in red colour. 

\end{itemize}

First, we have investigated the model for $z\in[0,2]$ and found that there is a clear deviation from $\omega_{dm} = 0$. In particular, we have found that its mean value is $\omega_{dm} =  -0.035$. Clearly, the obtained mean for $H_{0}$ is closer to the Planck Mission results than the zero value. On top of this, this slight hints at the possibility that the $H_{0}$ tension problem may be alleviated. On the other hand, the analysis shows that the value of $\omega_{dm}$ will increase when $z\in [0.2.5]$. But still, we have a clear deviation from cold dark matter. Moreover, we have learned that the mean of the free parameters $H_{0}$ and $\Omega_{dm}$  will increase, while the mean value of $\Omega_{b}$ will slightly decrease. The comparison of the results obtained hints here that, most likely, this model cannot fully solve the $H_{0}$ tension problem.  How could it change if we were able to measure the expansion rate up to redshift $z\ = 5$? What would be the result then? 

The answer can be seen from Table~\ref{tab:Table1} which indicates that the mean of the $H_{0}$ parameter will increase compared to the results obtained for the two previous redshift ranges. Moreover, the values of the parameters $\Omega_{dm}$ and $\Omega_{b}$ will decrease dramatically. We also found that $\omega_{dm} \neq 0$ and we have a hint that the deviation from cold dark matter will be preserved. The forecast results associated to the future possible measurements of the expansion rate data of our Universe when $z \in [0,5]$ read as follows: $H_{0} = 68.95 \pm 0.25$ km/s/Mpc,  $\Omega_{dm} = 0.271 \pm 0.003$, $\Omega_{b} = 0.00035^{+ 0.00038}_{-0.00022}$ and $\omega_{dm} = -0.028 \pm 0.001$. To end this subsection, we would like to mention the fact that it seems that the $H_{0}$ tension problem cannot be solved, even if there is a deviation from the usual cold dark matter case. Even if there is a tiny hint supporting the idea, it is actually not enough for the conclusion. Even more, we have observed that the high-redshift expansion rate observations may point to some dynamics of $\omega_{dm}$ and still indicate possible deviations from cold dark matter. 

\begin{figure}[t!]
 \begin{center}$
 \begin{array}{cccc}
\includegraphics[width=80 mm]{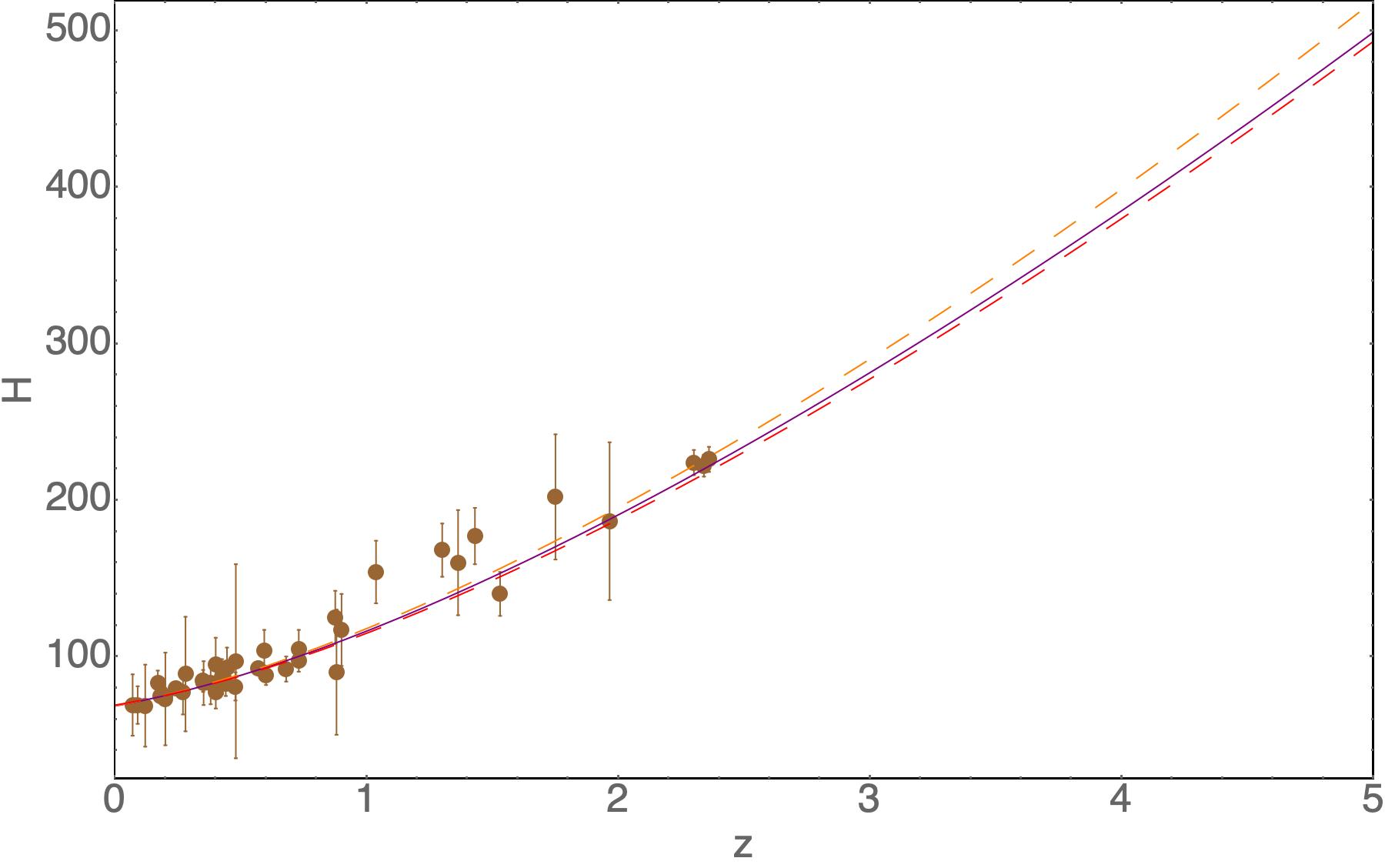}&&
\includegraphics[width=80 mm]{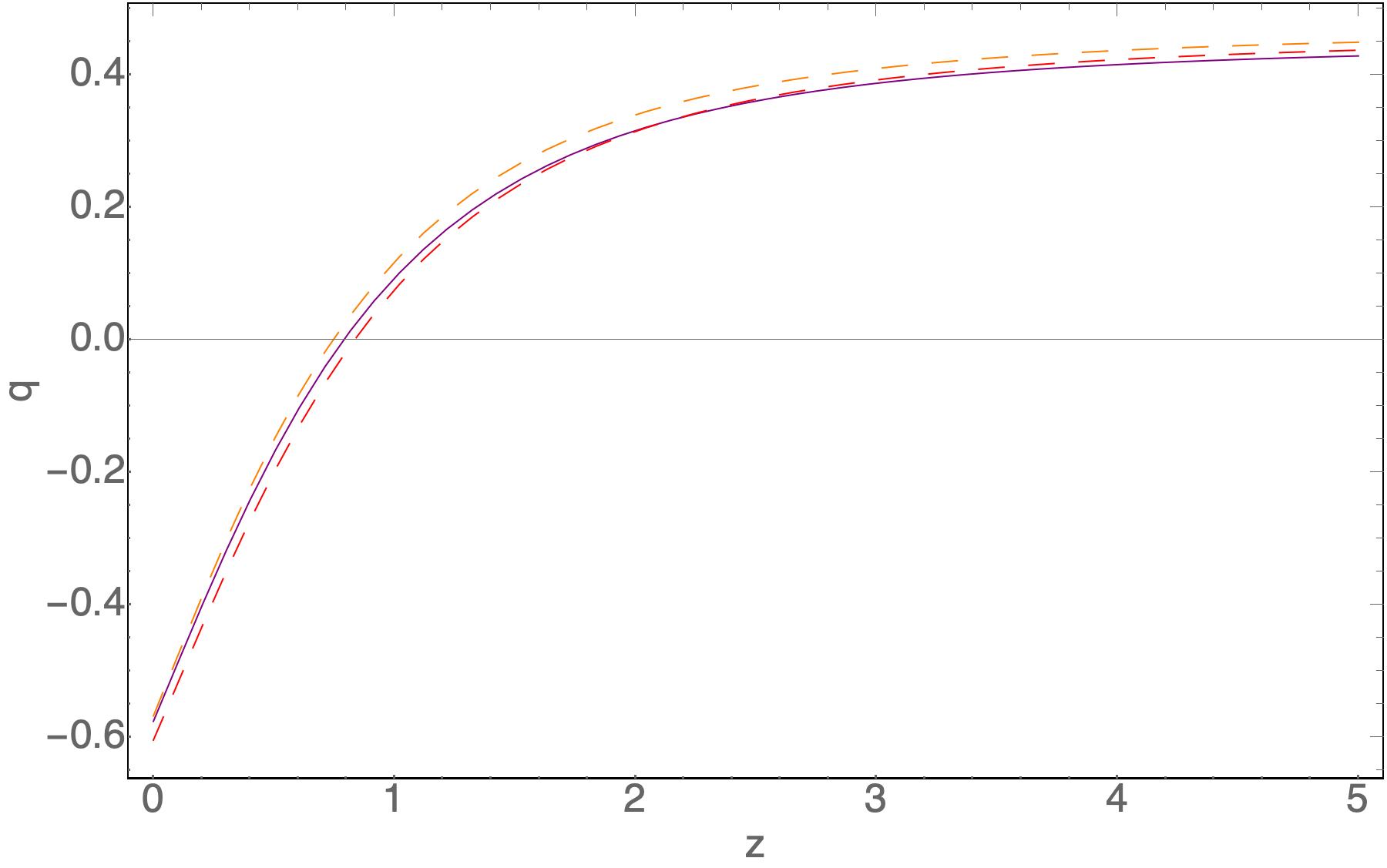}\\
 \end{array}$
 \end{center}
\caption{Graphic behaviour of the Hubble parameter compared with known $H(z)$ data is presented on the left-hand side. The purple curve is a plot of the Hubble parameter for the best fit values of the model parameters, when $z \in [0,2]$, the dashed orange curve corresponds to the case when $z \in [0,2.5]$, while the dashed red curve is for the case when $z \in [0,5]$. The red dots represent the known observational $H(z)$ data, which are the same as in Table~\ref{tab:Table0}. The right-hand side corresponds to the graphic behaviour of the deceleration parameter $q(z)$. The model is given by Eqs.~(\ref{eq:HXCDM}) and (\ref{eq:EXCDM_M1}). In both cases, only the best fit values for the model's free parameters obtained by the Bayesian Learning approach and depicted in Table~\ref{tab:Table1} have been used.}
 \label{fig:Fig0_2_a}
\end{figure}

The results discussed in the next subsection show that the assumption taken on the dark energy model can significantly affect the final results. The graphical behaviour of the expansion rate $H(z)$ and the deceleration parameter $q(z)$ corresponding to this second case can be found in Fig.~(\ref{fig:Fig0_2_a}). It certifies that we have performed a good learning strategy. Moreover, we see that the learned deviation for $\omega_{dm}$ is not only able to solve the $H_{0}$ tension, but it can also explain the BOSS experiment results for the expansion rate data at $z=2.34$ \cite{Alam:2016hwk}.
This is another interesting outcome of our work here, which requires future and more detailed study. On the other hand, the forecast results for $z\in [0,5]$ allow us to conclude that the high-redshift expansion rate data can validate that the learned $\omega_{dm}$ deviation can additionally be useful to explain the BOSS experiment result. Eventually,  we have learned very tight constraints for the model's free parameters, except for $\Omega_{b}$.   

\begin{table}
  \centering
    \begin{tabular}{ | c | c | c | c | c | c |  p{2cm} |}
    \hline
    
 Model 1 & $H_{0}$ & $\Omega_{dm}$ & $\Omega_{b}$ & $\omega_{dm}$ \\
      \hline
 
  when $z\in[0,2]$ & $68.42 \pm 0.24$ km/s/Mpc & $0.289 \pm 0.007$ & $0.0037 ^{+0.0032}_{+ 0.0025}$ &  $-0.035 \pm 0.005$ \\
          \hline
          
 when $z\in[0,2.5]$ & $68.54 \pm 0.24$ km/s/Mpc & $0.291 \pm 0.007$ & $0.0035 ^{+ 0.0036}_{-0.0024}$ &  $-0.022 \pm 0.005$ \\
          \hline
          
when $z\in[0,5]$  & $68.95 \pm 0.25$ km/s/Mpc& $0.271 \pm 0.003$ & $0.00035^{+ 0.00038}_{-0.00022}$ &  $-0.028 \pm 0.001$ \\

           \hline
 
     \end{tabular}
\caption{Best fit values and $1\sigma$ errors estimated for the $\Lambda$ + Baryonic Matter + Dark Matter (with $\omega_{dm} \neq 0$) model, when $z \in [0,2]$, $z \in [0,2.5]$ and $z \in [0,5]$, respectively. The results have been obtained from a Bayesian (Probabilistic) Machine Learning approach, where the generative process is based on Eq.~(\ref{eq:HXCDM}) and Eq.~(\ref{eq:EXCDM_M1}) using $H_{0} \in [64.0, 80.0]$,  $\Omega_{dm} \in [0.23, 0.4]$, $\Omega_{b} \in [0.0, 0.1]$ and $\omega_{dm} \in [-0.1, 0.1]$ flat priors. The analysis is based on 10 chains and in each chain, 10,000 "observational" data-sets from the model have been simulated or generated.}
  \label{tab:Table1}
\end{table}

\subsection{Model with $\omega_{dm} \neq 0$ dark matter and $\omega_{de} \neq -1$ dark energy}

\begin{table}
  \centering
    \begin{tabular}{ | c | c | c | c | c | c |  p{2cm} |}
    \hline
 
 Model 2 & $H_{0}$ & $\Omega_{dm}$ & $\Omega_{b}$ & $\omega_{dm}$ & $\omega_{de}$ \\
       \hline
 when $z\in[0,2]$ & $73.52 \pm 0.24$ km/s/Mpc & $0.271 \pm 0.012$ & $0.0036 ^{+0.0038}_{-0.0025}$ &  $-0.0505 \pm 0.0097$ & $ -1.049 \pm 0.044$\\
          \hline
when $z\in[0,2.5]$ & $73.47 \pm 0.23$ km/s/Mpc & $0.268 \pm 0.011$ & $0.0036 ^{+0.0041}_{-0.0022}$ & $-0.0354 \pm 0.0093$ & $-1.091 \pm 0.039$  \\
         \hline
 when $z\in[0,5]$  & $73.65 \pm 0.15$ km/s/Mpc & $0.254 \pm 0.008$ & $0.0069 \pm 0.0045$ & $-0.0752 \pm 0.0047$ & $ -1.064 \pm 0.009$  \\   
     \hline
    
    \end{tabular}
\caption{Best fit values and $1\sigma$ errors estimated for the X (Dark Energy with $\omega_{de} \neq -1$) + Baryonic Matter + Dark Matter (with $\omega_{dm} \neq 0$) model, when $z \in [0,2]$, $z \in [0,2.5]$ and $z \in [0,5]$, respectively. The results have been obtained from the Bayesian (Probabilistic) Machine Learning approach, where the generative based process is based on Eqs.~(\ref{eq:HXCDM}) and (\ref{eq:EXCDM}), using $H_{0} \in [64.0, 80.0]$,  $\Omega_{dm} \in [0.23, 0.4]$, $\Omega_{b} \in [0.0, 0.1]$, $\omega_{dm} \in [-0.1, 0.1]$, and $\omega_{de} \in [-1.5, -0.4]$ as flat priors. The analysis is based on 10 chains and in each chain, 10,000 "observational" data-sets from the model have been simulated or generated.}
  \label{tab:Table2}
\end{table} 

\begin{figure}[h!]
 \begin{center}$
 \begin{array}{cccc}
\includegraphics[width=120 mm]{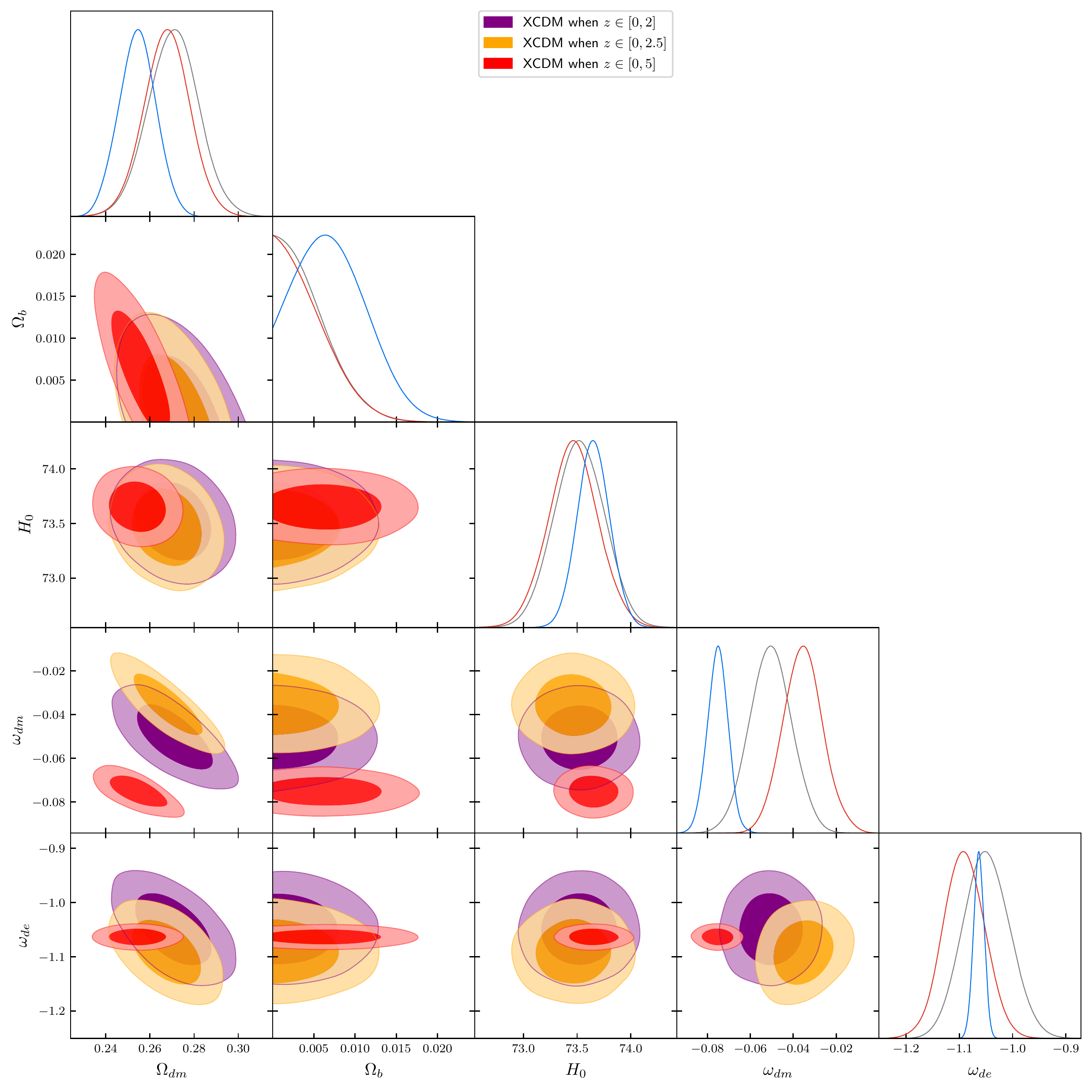}
 \end{array}$
 \end{center}
\caption{Contour maps of the model given by Eqs.~(\ref{eq:HXCDM}) and (\ref{eq:EXCDM}), for $z \in [0,2]$, $z \in [0,2.5]$, and $z \in [0,5]$, respectively. The best fit values of the model's parameters have been found to be $H_{0} = 73.52 \pm 0.24$ km/s/Mpc, $\Omega_{dm} = 0.271 \pm 0.012$, $\Omega_{b} = 0.0036 ^{+0.0038}_{-0.0025}$, $\omega_{dm} = -0.0505 \pm 0.0097$, and $\omega_{de} = -1.049 \pm 0.044$, when $z \in [0,2]$. While, when $z \in [0,2.5]$, the best fit values of the parameters have been found to be  $H_{0} = 73.47 \pm 0.23$ km/s/Mpc, $ \Omega_{dm} = 0.268 \pm 0.011$, $\Omega_{b} = 0.0036 ^{+0.0041}_{-0.0022}$, $\omega_{dm} = -0.0354 \pm 0.0093$, and $\omega_{de} = -1.091 \pm 0.039$. Finally, the best fit values of the model parameters have been found to be $H_{0} = 73.65 \pm 0.15$ km/s/Mpc, $ \Omega_{dm} = 0.254 \pm 0.008$, $\Omega_{b} = 0.0069 \pm 0.0045$, $\omega_{dm} = -0.0752 \pm 0.0047$ and $\omega_{de} = -1.064 \pm 0.009$, when $z \in [0,5]$. In all three cases, $H_{0} \in [64.0, 80.0]$,  $\Omega_{dm} \in [0.23, 0.4]$, $\Omega_{b} \in [0.0, 0.1]$, $\omega_{dm} \in [-0.1, 0.1]$ and $\omega_{de} \in [-1.5, -0.4]$ flat priors have been imposed during the generative process used to generate the "observational" data. The analysis is based on 10 chains and in each chain, 10,000 "observational" data-sets from the model have been simulated or generated.}
 \label{fig:Fig0_1_b}
\end{figure}

Our second model is the X (Dark Energy with $\omega_{de} \neq -1$) + Baryonic Matter + Dark Matter (with $\omega_{dm} \neq 0$) model where $\omega_{de}$ needs not be fixed to be the cosmological constant. Instead, it is a barotropic dark fluid with the constant equation of state parameter, which should be fitted. A detailed analysis of the model by using a Bayesian Learning approach shows that:
\begin{itemize}

\item The best fit values for the model's free parameters with $1\sigma$ error is $H_{0} = 73.52 \pm 0.24$ km/s/Mpc, $\Omega_{dm} = 0.271 \pm 0.012$, $\Omega_{b} = 0.0036 ^{+0.0038}_{-0.0025}$, $\omega_{dm} = -0.0505 \pm 0.0097$ and $\omega_{de} = -1.049 \pm 0.044$, when $z \in [0,2]$. The contour map is depicted in Fig.~(\ref{fig:Fig0_1_b}), in purple colour. 

\item As a second case, when $z \in [0,2.5]$, the most likely best fit values with $1\sigma$ error are: $H_{0} = 73.47 \pm 0.23$ km/s/Mpc, $ \Omega_{dm} = 0.268 \pm 0.011$, $\Omega_{b} = 0.0036 ^{+0.0041}_{-0.0022}$, $\omega_{dm} = -0.0354 \pm 0.0093$, and $\omega_{de} = -1.091 \pm 0.039$. The contour map is given in Fig.~(\ref{fig:Fig0_1_b}), in orange colour. 

\item Eventually, when $z \in [0,5]$, the most likely best fit values with $1\sigma$ error are: $H_{0} = 73.65 \pm 0.15$ km/s/Mpc, $ \Omega_{dm} = 0.254 \pm 0.008$, $\Omega_{b} = 0.0069 \pm 0.0045$, $\omega_{dm} = -0.0752 \pm 0.0047$, and $\omega_{de} = -1.064 \pm 0.009$. The contour map is given in Fig.~(\ref{fig:Fig0_1_b}), in red colour. 

\end{itemize}

We conclude from the results above that the $H_{0}$ tension can be solved and that there is a clear deviation from cold dark matter. Moreover, we see that this deviation is stronger than in previous case. Even more, the solution does not require any non-gravitational interaction between dark energy and dark matter. On the other hand, we observe that the deviation from cold dark matter has a clear redshift dependence and, moreover, if the expansion rate data for $z \in [0,5]$ would become available, then we should expect a very strong deviation. The results of the forecast for $z \in[0,5]$ gives us the following $H_{0} = 73.65 \pm 0.15$ km/s/Mpc, $ \Omega_{dm} = 0.254 \pm 0.008$, $\Omega_{b} = 0.0069 \pm 0.0045$, $\omega_{dm} = -0.0752 \pm 0.0047$ and $\omega_{de} = -1.064 \pm 0.009$ constraints. We notice that, for the higher redshift expansion rate data, the mean of $\Omega_{dm}$ will decrease significantly. For instance, if it is $\Omega_{dm} = 0.271$ for $z \in [0,2]$, then it will be $\Omega_{dm} = 0.268$ for $z \in [0,2.5]$. While we have obtained  $\Omega_{dm} = 0.254$, when $z\in [0,5]$.  On the other hand, the constraints on $\omega_{de}$ show that an increase in $\omega_{dm}$ will mean a decrease in $\omega_{de}$. Additionally, the  constraints obtained on $\omega_{de}$ in all three redshift ranges are in good agreement with existing observational data reported in the literature. Finally, we observe that the expansion rate data for $z\in [0,5]$ will indicate a significant change in $\Omega_{b}$~(see Table~\ref{tab:Table2}). 

To summarize our results, let us briefly mention where we started and what we have achieved. We have an $H_{0}$ tension and see that interacting dark energy models can solve it. If we have a closer look at introducing non-gravitational interaction between dark energy and dark matter, then it becomes clear that it affects the equation of state parameters of two energy sources. On the other hand, we have one extra degree of freedom, making the model softer for observational data. The question is why this should work and be useful for the $H_{0}$ tension problem? Can we have another look at the problem from a different perspective and figure out how it works? Here we worry only about the non-gravitational interaction since it is not clear why it should be used and how it appears in our Universe. Of course, we could argue that the observational data supports this idea. However, we should remember that it is true and workable when we have a specific dark energy model and cold dark matter, which is actually the only way to gauge it and does not give answers. 

Now, what we did in this paper. In particular, taking into account that the non-gravitational interaction affects the equation of state parameters of dark energy and dark matter models, we have applied Bayesian (Probabilistic) Machine Learning to study two cosmological models where, in each case, $\omega_{dm}$ is the free parameter that should be fitted. Surprisingly we have learned that $\omega_{dm} \neq 0$, both when dark energy is the usual cosmological constant and when it is given by another energy source, with $\omega_{de} \neq -1$. In both cases, we have learned the constraints on the free parameters and concluded that the $H_{0}$ tension problem is strongly dependent on the non-coldness of dark matter. Therefore, making dark matter colder for a given dark energy model without using non-gravitational interaction allows to solve (or at least alleviate) the $H_{0}$ tension problem. 

\begin{figure}[t!]
 \begin{center}$
 \begin{array}{cccc}
\includegraphics[width=80 mm]{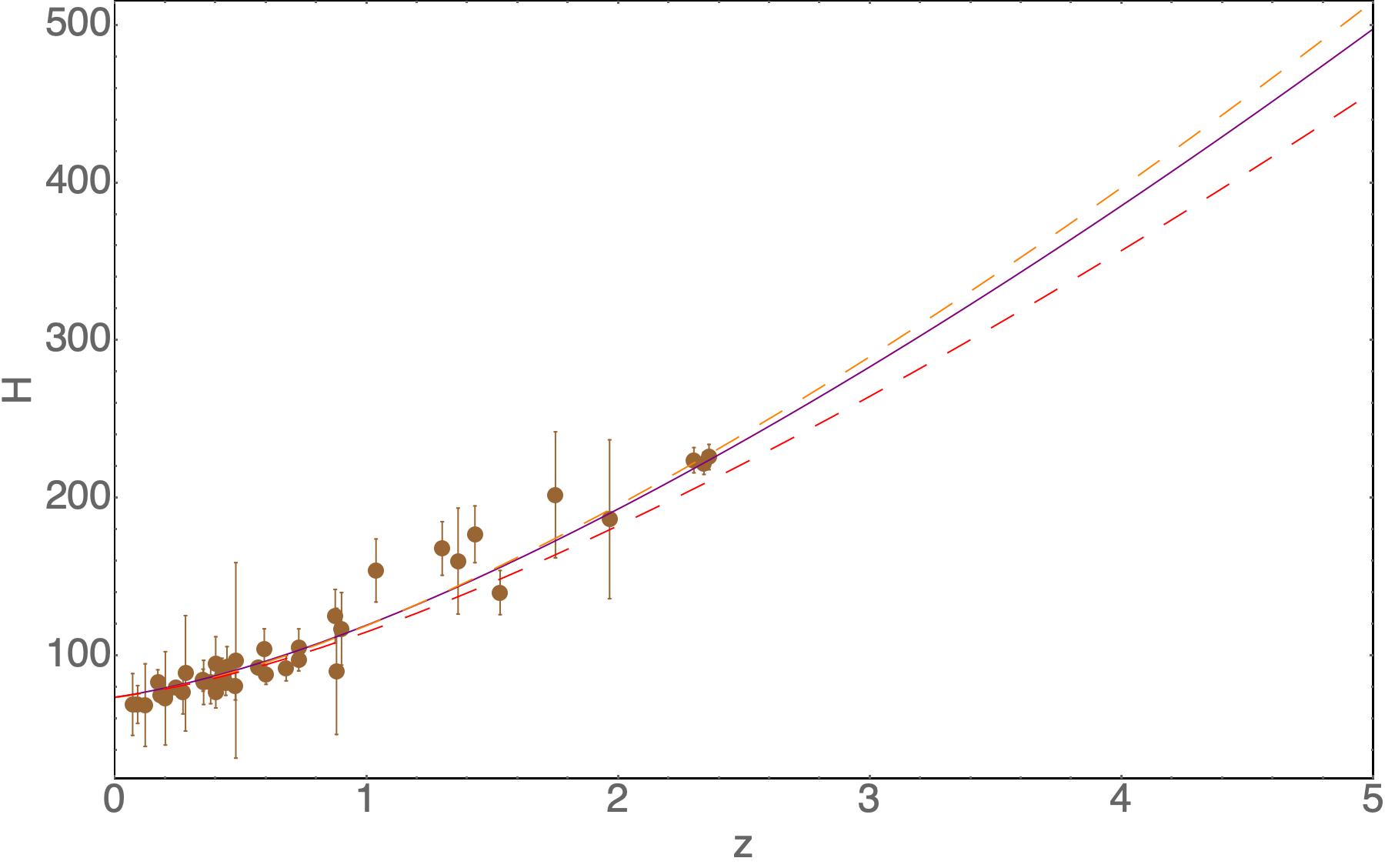}&&
\includegraphics[width=80 mm]{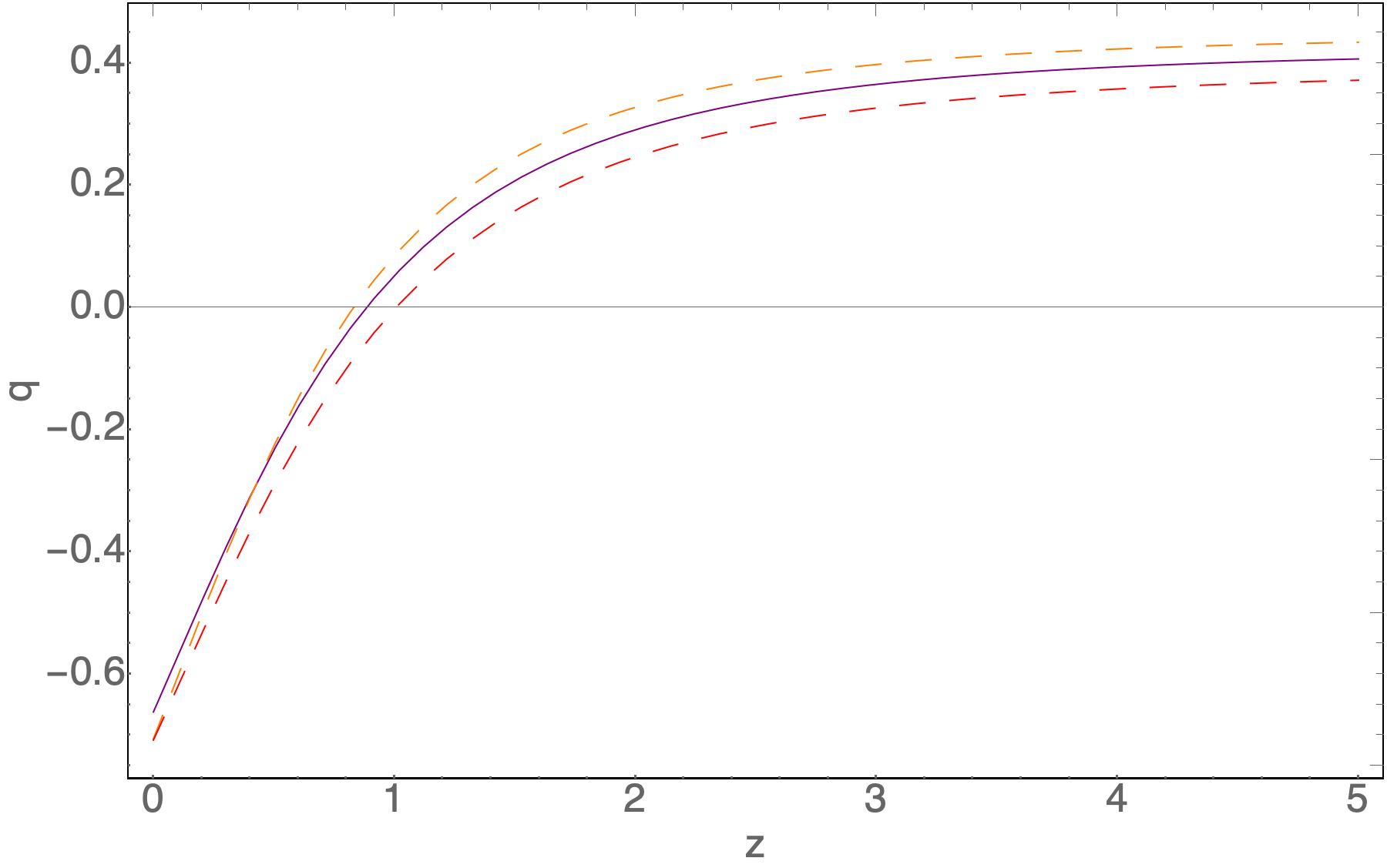}\\
 \end{array}$
 \end{center}
\caption{The graphic behavior of the Hubble parameter, in comparison with known $H(z)$ data, is presented on the left-hand side. The purple curve is a plot of the Hubble parameter for the best fit values of the model's parameters, when $z \in [0,2]$; the dashed orange curve corresponds to the case when $z \in [0,2.5]$; while the dashed red curve is for the case when $z \in [0,5]$. The red dots correspond to the known observational $H(z)$ data, and are the same as in Table~\ref{tab:Table0}. On the right-hand side, the graphic behavior of the deceleration parameter $q(z)$ is plotted. The model is given by Eqs.~(\ref{eq:HXCDM}) and  (\ref{eq:EXCDM}). In both cases, only the best fit values for the model's free parameters obtained by the Bayesian Learning approach and presented in Table~\ref{tab:Table2} have been used.}
 \label{fig:Fig0_2_b}
\end{figure}

Eventually, we would like to mention that the obtained constraints indicate how the interactions should be built. In particular, starting from the models not containing non-gravitational interaction, we found that, on the different cosmological scales, there is a clear and increasing deviation from the cold dark matter paradigm, which in the light of the $H_{0}$ tension problem could be confused to be an interaction between dark energy and dark matter. The graphic behavior of the expansion rate $H(z)$ and the deceleration parameter $q(z)$, corresponding to this case, can be found in Fig.~(\ref{fig:Fig0_2_b}). Similarly to the first case, here we see that very good learning has been achieved. Moreover, we see that the learned deviation for $\omega_{dm}$ is not only able to solve the $H_{0}$ tension, but it can also explain the BOSS experiment result of the expansion rate data at $z=2.34$~(the purple and orange curves in the left-hand side plot of Fig.~(\ref{fig:Fig0_2_b})). However, the forecast results for $z\in [0,5]$ hint that the high redshift expansion rate data can be in tension with the learned $\omega_{dm}$ deviation and could be not sufficient to explain the BOSS experiment result. Here, we should mention that this learned property can be used to impose the lowest bound on the $\omega_{dm}$ deviation. This is easy to see from a closer examination of the learned constraints presented in Table~\ref{tab:Table2}. Moreover, from Table~\ref{tab:Table2}, we see that we have learned very tight constraints for the model's free parameters, except $\Omega_{b}$. 

\section{Conclusions}\label{sec:conc}

Bayesian (Probabilistic) Machine Learning is one of the approaches used in modern science. It allows the use of the model itself to learn about its specific properties. It is a generative-based approach allowing us to explore the regimes in which we do not have physical data or where collecting them is a very costly process. It has been shown already that this procedure can be beneficial for other learning architectures by significantly reducing the associated costs. In this regard, the method can be very useful for predictive purposes too. In this paper, we use Bayesian (Probabilistic) Machine Learning to learn if and why interacting dark energy models should be useful in view of the $H_{0}$ tension problem. A closer look at interacting dark energy models indicates that the interaction does affect the equation of state parameters describing the two energy sources. However, why this interaction between dark energy and dark matter should exist? Could it be possible that we did not use proper parametrizations to describe the dark energy and dark matter components and that the interaction itself does only serve to correct this? Is this the reason why it seems to be quite useful to solve different cosmological problems? 

To better understand the situation, we decided to learn how precise our assumptions constrain the equation of state parameters of dark energy and cold dark matter. We started assuming that the dark matter $\omega_{dm}$ equation of state parameter is one of the free parameters to be fitted in the context of the $\Lambda$ + Baryonic Matter + Dark Matter (with $\omega_{dm} \neq 0$) and  X (Dark Energy with $\omega_{de} \neq -1$) + Baryonic Matter + Dark Matter (with $\omega_{dm} \neq 0$) models. The constraints have been obtained for three growing redshift ranges, $z\in [0,2]$, $z\in [0,2.5]$ and $z\in [0,5]$, respectively. The first two redshift ranges are chosen to validate our fit results since we have very good quality observations of the expansion rate $H(z)$ in these cases. On the other hand, considering the extended redshift range, $z\in [0,5]$, is for forecasting reasons. The full validation of the results will have to wait for available observations of higher-redshift $H(z)$ data. 

Surprisingly, we learned that $\omega_{dm} \neq 0$, both when dark energy is the cosmological constant and an energy source with $\omega_{de} \neq -1$. We clearly see a deviation from the $\Lambda$CDM standard model. The learning results based on the generated expansion rate data from the background dynamics of each model show that, on the cosmological scales, there is a deviation from the cold dark matter hypothesis. The results obtained explicitly demonstrate that the deviation from cold dark matter can effectively be a solution for the $H_{0}$ tension problem. In particular, we learned that, by only making dark matter colder for a given dark energy model different from the cosmological constant without using a non-gravitational interaction, one might get a solution to the $H_{0}$ tension problem. Moreover, our results hint at why the parameters of the interaction term should be small, as often discussed in the literature. 

On the other hand, we saw that the learned deviation for $\omega_{dm}$ for the $\Lambda$ + Baryonic Matter + Dark Matter (with $\omega_{dm} \neq 0$) can not solve the $H_{0}$ tension, but it is able to explain the BOSS experiment result of the expansion rate data at $z=2.34$. It should be mentioned that the forecast results for $z\in [0,5]$ allow us to conclude that the high redshift expansion rate data can validate that the learned $\omega_{dm}$ deviation still may be useful to explain the BOSS experiment result. Contrary to the first case, the analysis of the X (Dark Energy with $\omega_{de} \neq -1$) + Baryonic Matter + Dark Matter (with $\omega_{dm} \neq 0$) model shows that the learned deviation for $\omega_{dm}$ does the job. However, the forecast results for $z\in [0,5]$ hint that the high-redshift expansion rate data could be in tension with the learned $\omega_{dm}$ deviation and that they might not be sufficient to explain the BOSS experiment result. Here, we should mention that this learned property can impose the lowest bound on the $\omega_{dm}$ deviation. This is easy to see, particularly from a closer examination of the learned constraints presented in Table~\ref{tab:Table2}. The results of Table~\ref{tab:Table1} and those of  Table~\ref{tab:Table2} indicate that we can learn very tight constraints on the free parameters, except for $\Omega_{b}$.

Finally, we should duly stress that, while playing with the equation of state parameter of dark matter, we have achieved some remarkable results indicating a clear deviation from the cold dark matter paradigm. The last result could be applied to solve the $H_{0}$ tension indicating new physics, which should be explored further. They could be applied to explain the BOSS experimental result, too. In the present work, we have restricted ourselves to two specific dark energy models only. However, other viable dark energy models are available,  and it will be most interesting to investigate them to learn more about the $\omega_{dm}$ deviation. Therefore, we plan to extend our work to consider more complicated cases involving other dark energy models. We also plan to study the structure formation problem, which is key to understanding how a deviation from cold dark matter will affect structure formation and stability. Of course, understanding to which extend the learned deviation can be useful to overcome various cosmological problems (as a replacement for involving interacting dark energy models) is a crucial issue that we plan to address in new situations. Progress on the mentioned issues will be reported in forthcoming papers.  To end, we would like to recall that the whole analysis presented in this paper has been done employing a PyMC3 python-based framework, where the expansion rate data has been generated and used during the learning process. Only after that, the learned constraints on the model's free parameters have been used to plot the graphical behaviour of the $H(z)$ parameter for each model to compare it with the available observational data.

\section*{Acknowledgements}

This work has been partially supported by MCI/AEI/FEDER, European Union, projects FIS2016-76363-P and PID2019-104397GB-I00, by AGAUR (Catalan Government), project 2017-SGR-247, by the Polish National Science Centre (NCN) under the Grant Agreement 2020/37/B/ST2/02371 and the research activities co-financed by the funds granted under the Research Excellence Initiative of the University of Silesia in Katowice.
 


\end{document}